\documentclass[aps,prl,reprint,10pt]{revtex4-1}

\usepackage{graphicx} 
\usepackage{comment}
\pdfoutput=1
\begin{document}
\title{Scale of left-right symmetry}

\author{Ravi Kuchimanchi}
\email{ravikuchimanchi1@gmail.com}
\begin{abstract}
Unlike  the standard model where neutrino masses can be made arbitrarily small, we find in the minimal left-right symmetric model  that Dirac type Yukawa coupling $h_D \sim 10^{-4.2}$ for  $\nu_\tau$ is generated from charged fermion Yukawa couplings through one loop Renormalization Group (RG) running. 
Using the seesaw relation this implies that the natural scale for right-handed neutrino's Majorana mass (seesaw scale) $M_N = f v_R \gtrsim 2000 TeV$.  If we take 
the $SU(2)_R \times U(1)_{B-L}$ breaking  scale $v_R$
to be below $50 TeV$, so that it can be probed by LHC and a future collider, then the large $h_D$ generated on $RG$ running
increases the mass $m_\nu$ of the third generation light neutrino and severely suppresses the PMNS mixing angle $\theta_{23}$. 
We discuss the tuning needed for seesaw scales that can be probed by colliders and find the parameter space that can be tested by neutrino experiments for higher scales up to $10^{15} GeV$. 
\end{abstract}

\maketitle



 The well known left-right symmetric (LR) model~\cite{PhysRevD.10.275,PhysRevD.11.566,Senjanovic:1975rk} restores parity between left and right in nature's Lagrangian 
and is one of the most elegant theories beyond the standard model. Historically, the Standard Model predicted that neutrinos are massless. This has been its biggest failure so far. 
In the minimal LR model, corresponding to the left-handed neutrino, there is a parity symmetric right-handed neutrino. Therefore the neutrinos \emph{can} have Dirac masses ($m_D$), like any other fermions, but \emph{do} they?  

If the neutrino is a Majorana particle, $m_D$ contributes to the light neutrino mass via the seesaw relation $m_\nu = m_D^2/M_N$. 
The question that motivates this work is, can $m_D$ naturally be zero in the minimal LR model? \emph{If not, then we should be able to calculate the natural minimal value for it.} 

With this value for $m_D$, and the observed light neutrino mass $m_\nu \sim \sqrt{\Delta {m^2_{32}}} = 0.05 eV$, we can use the seesaw relation to get a lower bound on the right handed neutrino's Majorana mass (seesaw scale) $M_N$, which  also serves as a bound on the natural scale of left-right symmetry $v_R$. 

In the standard model (with $\nu_R$), if the Lagrangian is symmetric under $\nu_R \rightarrow -\nu_R$, then $m_D = h_D v_{wk}$ vanishes. The smaller the  $h_D$, the lesser is this symmetry broken. Thus the idea of a $TeV$ scale seesaw, which has at its core naturally small Dirac Yukawa $h_D \sim 10^{-6}$ for the neutrino, is fueled by standard model thinking. 

In the minimal LR model, corresponding to the left handed lepton doublet $L_{3_L} \equiv (\nu_\tau, \tau^-)^T_L$ there is a right-handed $SU(2)_R$ doublet $L_{3_R} \equiv (\nu_\tau, \tau^-)^T_R$. Under the discrete symmetry $\nu_{\tau_R} \rightarrow -\nu_{\tau_R}$,   the Dirac Yukawa coupling of $\tau^-$ will also vanish.  It is easy to see that one of the top, bottom or tau Yukawas must vanish if a symmetry sets $h_D$ to zero. Therefore under renormalization group running, the neutrino Dirac Yukawa terms will get generated from charged fermion Yukawas, even if we take $h_D$ to be zero at a cut-off scale such as the Planck scale. 

In this work we obtain the one loop Renormalization Group Equation for the leptonic Yukawa coupling matrix $h^\ell$ of the LR Model, and find that $h^\ell_{33}\sim 10^{-4.2}$ is generated from the known Yukawa couplings of third generation charged fermions. 
$h^\ell_{33}$ is the Yukawa coupling with which  $\nu_\tau$ interacts with the neutral component of the first standard model Higgs doublet in the bidoublet of the LR Model. Thus it contributes directly to $h_D$ and for  technical naturalness $h_D \gtrsim 10^{-4.2}$.  Using the seesaw relation we obtain from the above, a bound on the seesaw scale $M_N \gtrsim 2000 TeV$. 

For seesaw scales $\sim 1-50 TeV$  that are accessible at LHC or a future $100$ TeV collider,  we would need to tune $h_D$ by $1\%-10\%$ so that it can have smaller values $\sim 10^{-5.5}$. This can be done by tuning the parameters of the Higgs potential so that on RGE running,  an appropriate VEV is generated for the second standard model Higgs doublet (in the bi-doublet), that provides a canceling mass to $m_D$.   All work in $TeV$ scale LR phenomenology, has so far assumed that no tuning is needed for $h_D \sim 10^{-5.5}$.  

The seriousness of the naturalness issue can be seen from bottom-up.  We show that if we just take the neutrino Yukawas to be small so as to have a $TeV$ scale seesaw as is currently done, then  the larger values of $h^\ell_{33}$ generated on RGE running to just a few magnitudes higher scale, will  increase the third generation light neutrino mass $m_\nu$  and  suppress the PMNS mixing angle $\theta_{23}$ so that it is no longer large.  That is, the large mixing angle $\theta_{23}$ is unstable under RGE running. Moreover, even if we tune and generate a canceling contribution to $m_D$ (and therefore to $m_\nu$) at a specific renormalization scale,  the suppression of $\theta_{23}$ may still persist at other scales.       

This work is important also because it is quite general.  Since we only need to consider the right-handed neutrino's Majorana mass $M_N$  in our calculation, it does not depend on the mass generated from Majorana Yukawa couplings of the left-handed neutrino.  Therefore our result applies even if the discrete parity symmetry $P$ is broken at a higher scale than the gauge symmetry breaking scale $v_R$, so that the left and right sectors have different parameters (such as discussed in section 3 of~\cite{Mohapatra2016423}).  
A similar analysis can be done for the supersymmetric case, as even in the minimal SUSY LR model~\cite{Kuchimanchi:1993jg} there is no symmetry that can set \emph{only} the neutrino couplings to zero.  

We also provide a useful figure that shows the natural and fine tuned regions of the left-right symmetric model, and the reach of various experiments -- both collider and those that probe the leptonic CP phase, and thereby captures the big picture for testing the minimal left-right symmetric model in the entire range of seesaw scale from $TeV$  to $10^{15} GeV$.

\vskip0.1in

\noindent \emph{Generation of Dirac mass and seesaw scale --}

We consider the well-known minimal Left-Right symmetric model~\cite{PhysRevD.10.275,PhysRevD.11.566,Senjanovic:1975rk} based on $G_{LR} \equiv SU(3)_c \times SU(2)_L \times SU(2)_R \times U(1)_{B-L} \times P$, with scalar triplets $\Delta_R$ (1,1,3,2) and $\Delta_L$ (1,3,1,2), and bi-doublet $\phi$ (1,2,2,0). Under parity ($P$),  the space-time coordinates $(x,t) \rightarrow (-x,t), \ \phi \rightarrow \phi^\dagger$ and subscripts $L \leftrightarrow R$ for all other fields (see for example~\cite{Duka:1999uc}).   The scalar fields have the form
\begin{equation}
\begin{array}{ccc}
\phi = \left(\begin{array}{cc}
\phi^o_1 & \phi^+_2 \\
\phi^-_1 & \phi^o_2
\end{array}
\right), 
 &
\Delta_{L,R} = \left(\begin{array}{cc}
\delta^+_{L,R} / \sqrt{2} & \delta^{++}_{L,R} \\
\delta^o_{L,R} & - \delta^+_{L,R} /\sqrt{2}
\end{array}
\right), 
\end{array}
\label{eq:fields}
\end{equation}
 Note that there are two standard model Higgs doublets in the bidoublet, and are indicated by the subscripts $1$ and $2$ in matrix elements of $\phi$. The second standard model Higgs doublet is naturally heavy and is at the $SU(2)_R \times U(1)_{B-L}$ breaking scale $\left<\delta^o_R\right> =v_R$. 

When the bidoublet $\phi$ picks up weak-scale VEV diag$\{k_1,k_2\}$ with $v_{wk}^2 = k_1^2 + k_2^2 = 174^2 GeV^2$, the quarks ($Q_i$) and leptons ($L_i$) pick up Dirac type mass contributions through the Yukawa terms
\begin{equation}
\bar{Q}_{i_L} (h_{ij}\phi +\tilde{h}_{ij} \tilde{\phi})Q_{j_R} +\bar{L}_{i_L} (h^\ell_{ij} \phi +\tilde{h}^\ell_{ij} \tilde{\phi})L_{i_R} + h.c.
\label{eq:diracyukawa}
\end{equation}
where $\tilde{\phi} = \tau_2 \phi^\star \tau_2$.   $h,\tilde{h},h^\ell$ and  $\tilde{h}^\ell$ are $3\times 3$ Yukawa coupling matrices in the generation space, and are Hermitian due to $P$. Without loss of generality we take $|k_1| > |k_2|$ and the VEVs to be real.

Barring cancellations, the only way to avoid Dirac masses for the neutrinos is to have the Yukawa matrix $h^\ell = 0$ \emph{and} the VEV $k_2 = 0$. Even if we consider only the third generation, there is no symmetry of the Yukawa terms in eqn.~(\ref{eq:diracyukawa}) that can achieve this, without also setting one of either top, bottom or tau masses to zero.  

For example, consider a $Z_4$ symmetry under which $L_{i_R} \rightarrow iL_{i_R}$ and $\phi \rightarrow i \phi$ that can set $h^{\ell}$ to zero, but this is broken by $h$ if under $Z_4$, $Q_{i_R} \rightarrow i Q_{i_R}$ (or by $\tilde{h}$ if $Q_{i_R} \rightarrow -iQ_{i_R}$).  Note that both $h$ and $\tilde{h}$ must be non-zero to provide masses to the top and the bottom quarks since we also require that $k_2=0$.

Note also that a $Z_2$ symmetry under which $L_{i_R} \rightarrow -L_{i_R}$ will likewise be broken by $\tilde{h}^\ell$ which gives the tau a mass in the absence of $k_2$.

\begin{figure}
\begin{center}
\includegraphics[height=3cm]{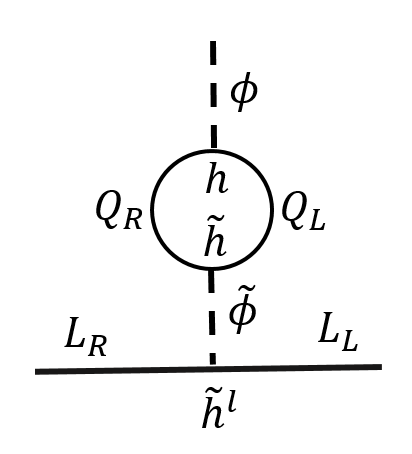}
\end{center}
\caption{One loop contribution to renormalization group running of $h^\ell$ calculated in eqn~(\ref{eq:RGE1}) using Yukawa couplings from eqn.~(\ref{eq:diracyukawa}) and method in~\cite{Cheng:1973nv}.}   
\label{rgefig}
\end{figure}

Therefore even if we set $h^\ell$ to zero at some scale, it gets generated via the following term of the one-loop Renormalization Group Equation that violates both $Z_4$ and $Z_2$ (see Fig~\ref{rgefig}).

\begin{equation}
\frac{dh^\ell}{d(ln\mu)} = \left[\frac{6}{16 \pi^2}\right]\tilde{h}^\ell Tr\left(h\tilde{h}\right)
\label{eq:RGE1}
\end{equation} 

The factor 6 comes from 3 colors of top and bottom running in the loop.  In the RHS of the above we have ignored the remaining terms as they conserve $Z_4$ and they all vanish in the limit $h^\ell \rightarrow 0$. Note also that the above term is analogous to the second term in equation (409) of the two Higgs doublet model in reference~\cite{Branco:2011iw} that generates the Yukawa couplings of the charged leptons with one standard model Higgs doublet from their Yukawa couplings with the other Higgs doublet. 

Considering only the larger couplings of the third generation, we can express the RHS of eqn.~(\ref{eq:RGE1}) in terms of the known top, bottom and tau Yukawas by identifying: 
\begin{equation}
h_{33} = h_t, \ \tilde{h}_{33} = h_b, \ \tilde{h}^\ell_{33} = h_\tau
\end{equation}
 
 For the running top, bottom and tau Yukawas, we use the one loop Renormalization Group Equations for gauge and Yukawa couplings of the LR model that are worked out in reference~\cite{Rothstein:1990qx} and take their values from refs~\cite{Buttazzo:2013uya},~\cite{Bednyakov:2016onn} at renormalization scale $\mu = 10 TeV$ to be $h_t=0.77$, $h_b=0.013$, $h_\tau=0.01$, $g_1=0.47$, $g_2=0.62$ and $g_3=0.95$. Note that reference \cite{Rothstein:1990qx} does not have the term in eqn~(\ref{eq:RGE1}) as it appears to have overlooked it.

We now calculate the value of $h^\ell$ obtained at the $TeV$ scale, if we take $h^\ell = 0$ at the Planck scale $M_{Pl}  \sim 10^{18} GeV$. The $h^\ell_{33}$ thus generated at the TeV scale using equation~(\ref{eq:RGE1}) and running values of the charged fermion Yukawas is:
\begin{equation}
-h^\ell_{33} = 6.1 \times 10^{-5} = 10^{-4.2}
\end{equation} 
\begin{figure}[t]
\begin{center}
\includegraphics[width=8.8cm]{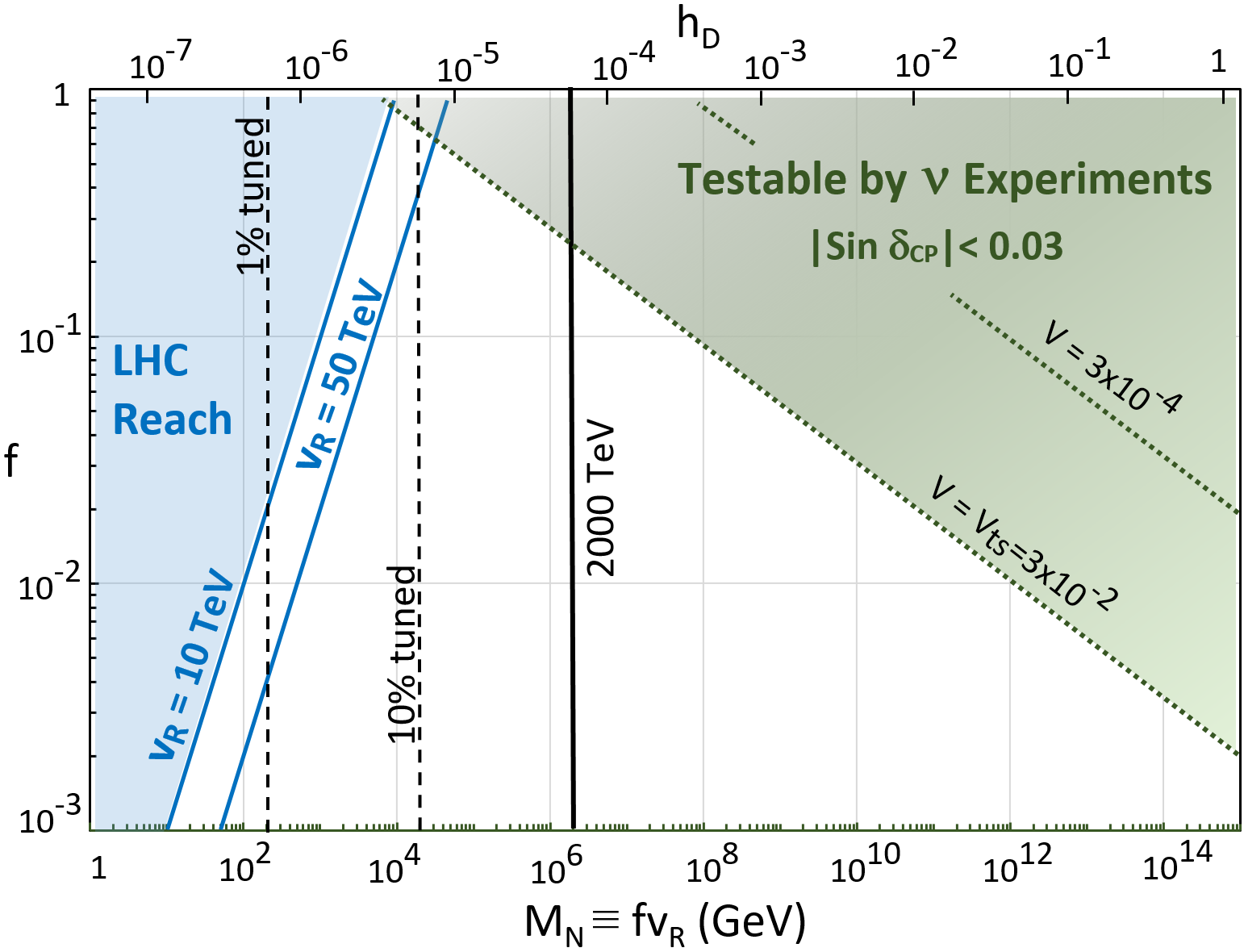}
\end{center}
\caption{Figure shows the natural and tuned regions of the parameter space of the minimal left-right symmetric model and the reach of various experiments.   
The natural value $h_D\sim10^{-4.2}$ generated by RGE running corresponds to $M_N\sim2000 TeV$ and the regions to the right of this line do not require any tuning.  The vertical dashed lines correspond to $10\%$ and $1\%$ tuning of $h_D$.
The y-axis is the Majorana Yukawa coupling $f$ of $\nu_{\tau_R}$.  The LHC reach is shown to be to the left of $v_R = 10 TeV$ line,   
and a $100 TeV$ collider can probe $v_R = 50 TeV$.  On these two lines, as can be seen in the figure, regions with smaller values of $f$ have smaller $h_D$ and are more tuned.
A significant part of the natural region of parameter space is testable by neutrino experiments that aim to measure the leptonic CP phase $\delta_{CP}$. For mixing angle $V = V_{ts} = 3 \times 10^{-2}$ between the second and third generation leptonic Dirac Yukawas, and Type 2 seesaw dominance,  $|\sin \delta_{CP}|$ must be less than $0.03$ so that its contribution to the strong CP phase $\bar{\theta}$ is $\leq 10^{-10}$. This is  the green shaded region above the lower slanting dotted line.  For much smaller mixing $V = 3 \times 10^{-4} $,  it is the smaller region above the higher slanting dotted line. For Type 1 seesaw dominance, the green shaded region will shift up along the $f$ axis by a factor of about 10, as discussed in the text. }   
\label{fig1}
\end{figure}

The above is the natural value for the Dirac Yukawa coupling of the neutrino generated by RGE running in the minimal Left Right symmetric model.    $h^\ell$ contributes $h^\ell_{33} k_1/\sqrt{2}$ to the Dirac mass $m_D$ of the neutrino.  Using the seesaw formula $m_\nu = m_D^2/M_N$ we find for the natural value of the seesaw scale
\begin{equation}
M_N \gtrsim (10^{-4.2} \times 174 GeV)^2 /(0.05 eV) = 2000 TeV
\end{equation}
where we have taken the observed light neutrino mass $m_\nu = 0.05 eV$ and $k_1 \approx 174 GeV$.  

For a $TeV$ scale seesaw, there must be a canceling contribution to $m_D$, in effect a tuning of the effective Dirac Yukawa coupling of the neutrino so that it is lowered from $10^{-4.2}$ to $10^{-6}$. 
 

Since $m_D = h^\ell_{33} k_1 + \tilde{h}^\ell_{33} k_2 \equiv h_D v_{wk}$ (with $\tilde{h}^\ell_{33}= h_\tau$), one way to cancel it is by fine-tuning the VEV $k_2$ of the second standard model doublet in the bi-doublet.   $k_2$ is obtained by minimizing with respect to $k_2$ the Higgs potential terms that have the form $V = \mu_{eff}^2 k_1 k_2 + \alpha_3 v^2_R k^2_2$,
where as is well known, $\alpha_3$ is the dimensionless parameter of a quartic term of the Higgs potential of LR model using the notation of reference \cite{Duka:1999uc}, and $\alpha_3 v_R^2$ is the mass of the second SM Higgs doublet. $\mu_{eff}^2k_1k_2$ is obtained from quadratic and quartic terms of the Higgs potential that contain $Tr(\tilde{\phi}^\dagger\phi)$, evaluated with VEV $\left<\delta^o_R\right> = v_R$.  By an appropriate choice of the parameters and their RGE running, we can generate $k_2$ so that 
its contribution  to $m_D$ cancels to about a $10\%$ to $1\%$ level, the contribution that was obtained due to generation of $h^\ell_{33}$.  Figure~\ref{fig1} shows the tuning needed for different values of $h_D$ (or equivalently $M_N$). On the y-axis is the Yukawa coupling $f\equiv f_{33}$ of the Majorana term $if_{33} L_{3_R}^T\tau_2 \Delta_R L_{3_R}$.

\vspace{0.2cm}

\noindent  \emph{How serious is the problem?}

 Supposing we just take $h_{33}^\ell = h_D= 3 \times 10^{-6}$ and $f = 0.6$ at low energies,  so that we have $v_R \sim 10 TeV$ and $M_N \sim 6 TeV$, then what is the problem?
 
Using equation~(\ref{eq:RGE1}), we find that larger values of $h_{33}^{\ell} \sim  10^{-5}$ and higher get generated through RGE running of just few orders of magnitude from $10 TeV$ to $100-10000 TeV$ scale. As shown in Table~\ref{table:2}, since $m_\nu = (h_{33}^\ell v_{wk})^2 / M_N $, it rises  with the increasing $h_{33}^\ell$.  Note that in a basis in which charged lepton mass matrix (or $\tilde{h}^\ell$ ) is diagonal, equation~(\ref{eq:RGE1}) generates only a diagonal contribution to $h^\ell$, and not the mass terms in the mixing.  This means that the PMNS mixing angle $\tan^2\theta_{23}$ between second  and third generation leptons decreases as $\sim m_\nu\{v_R\}/m_\nu\{\mu\}$ and we lose large neutrino mixing on RGE running. 

\begin{figure}[t]
\begin{center}
\includegraphics[width=8.5cm]{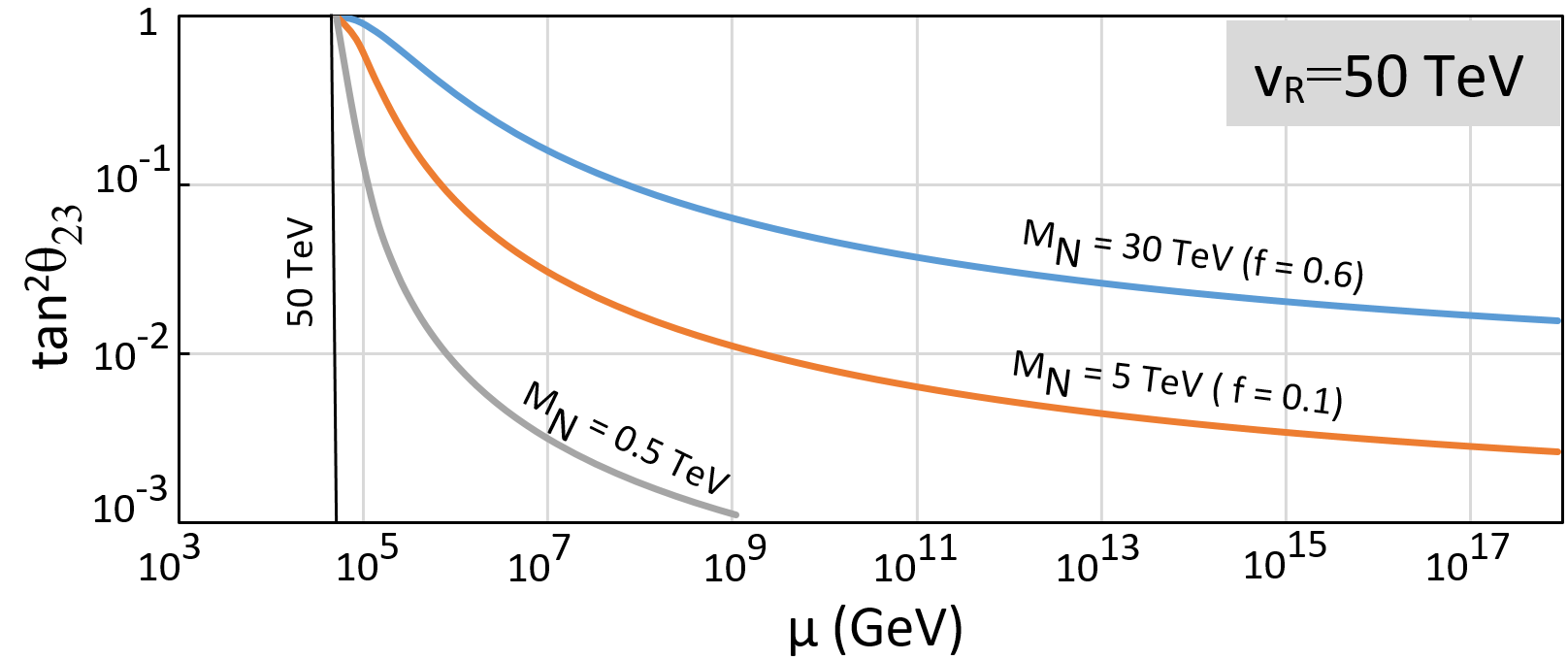}
\end{center}
\caption{PMNS mixing angle $\tan^2\theta_{23}$ 
is suppressed on RGE running even with $v_R = 50 TeV$ that can be reached by a $100TeV$ collider. The suppression is more severe for smaller values of $f$ (or $M_N = fv_R$). We have taken $f<0.7$ so that it remains perturbative. 
}   
\label{figtan}
\end{figure}

From Figure~\ref{figtan} we can see that the suppression of $\tan^2{\theta_{23}}$ is a concern even for the case of $v_R \sim 50 TeV$ that may be probed at a future $100 TeV$ collider.  

  By tuning we can cancel the contribution to $m_\nu$ at some chosen renormalization scale, so that there is no suppression of $\theta_{23}$ at that scale.   However at few magnitudes lower or higher scales the suppression may persist. Tuning may be more effective for higher values of $M_N \sim 30 TeV$ that may be probed by a $100 TeV$ collider. 


\begin{table}[t]
$
\begin{array}{cc|ccccc}
 \hline
 \mu & 10 TeV & 100 TeV & 1000 TeV & 10^7GeV & 10^{18} GeV \\ 
 \hline\hline
 m_\nu\{\mu\} & 0.05 eV & 0.3 eV & 1.1 eV & 2.4 eV & 18.7 \\ 
 \hline
\tan^2\theta_{23} & 1 & 0.16 &0.04 & 0.02  & 0.003  \\
\hline
\end{array}
$
\caption{If $v_R = 10TeV, M_N = fv_R = 6TeV$ so that the physics is within LHC reach, then the $h^\ell_{33}$ generated from eq~(\ref{eq:RGE1}) on RGE running to renormalization scale $\mu$, increases the mass $m_\nu\{\mu\}$ of the third generation light neutrino.  Therefore the PMNS mixing angle $\tan^2\theta_{23} \sim m_\nu\{v_R\}/m_\nu\{\mu\}$ gets suppressed.   
}
\label{table:2}
\end{table}

\vspace{0.2cm}
\noindent \emph{Testing left-right symmetry}

An important question to ask is, can we test the natural regions of the left-right symmetric model?  Curiously naturalness itself comes to the rescue. 

In the left-right symmetric model, the strong CP phase $\bar{\theta}$ generated on parity breaking can be calculated in terms of CP phases in the Higgs potential and Yukawa couplings.  The VEV $\left<\delta^o_R\right> \sim v_R$ that gives large Majorana masses to the right handed neutrinos breaks $P$ in the Higgs and leptonic sectors, and therefore if $CP$ violation is also present in these sectors it can easily generate a strong $CP$ phase. As shown in~\cite{Kuchimanchi:2014ota},  a large $\bar{\theta}$  is spontaneously generated at the tree-level from a $CP$ violating quartic term in the Higgs potential, and the one loop correction to it from $CP$ phases in the leptonic Yukawa sector.  Therefore in likely regions of parameter space of the LR model (without axions), naturalness  requires that the CP violation in Higgs and leptonic sectors be negligibly small, so that $\bar{\theta} \leq 10^{-10}$.  

Quantitatively, the Dirac CP phase $\delta_{CP}$ in the leptonic sector is constrained by the following relation obtained from equation~(6) of ref~\cite{Kuchimanchi:2014ota} written in a basis where $h^\ell$ is diagonal:   
\begin{equation}
|f_{33} f_{32}\tilde{h}^\ell_{23}h^\ell_{33} \sin\delta_{CP}| \leq 0.3\times \bar{\theta}
\label{eq:lepcp}
\end{equation} 

To plot the above region in Fig~\ref{fig1} we first re-write the above in terms of $f$ and $h_D$ by noting that $f_{32} \sim f_{33} \equiv f$ for Type-2 seesaw dominance, $h^\ell_{33} \sim h_D$,  $\tilde{h}^\ell_{23} \sim  V h_\tau$ and $\bar{\theta} \sim 10^{-10}$.   For the mixing $V$ in the leptonic Dirac Yukawa sector we  consider two cases: $V\sim V_{ts} = 3\times 10^{-2}$ like in the quark sector, and the much smaller $V = 3 \times 10^{-4}$. Eqn.~(\ref{eq:lepcp}) now implies that  $|\sin\delta_{CP}| \leq 0.03$ in regions \emph{above the respective dotted lines} for these two cases, as shown in Fig~\ref{fig1}.  


For a Type-1 seesaw dominance, $f_{32} \sim f_{33}/100$, and the region testable by neutrino experiments in Fig~\ref{fig1} will shift upwards along the $f-$axis by the factor $\sqrt{100}$.

$\delta_{CP}$ is being measured by neutrino experiments, and $CP$ conserving value $\sin \delta_{CP} =0$  is within  $70\%$ CL ($1.05 \sigma$) of current global fits~\cite{Esteban:2016qun}.

\vspace{0.2cm}

\noindent    \emph{Conclusion} 

In this work we did a basic calculation of the natural scale above which left-right symmetry may be found,  in terms of the the top, bottom and tau Yukawa couplings and the observed light neutrino mass $\sqrt{\Delta m_{32}^2}$.      We find that $h_D \sim 10^{-4.2}$ for $\nu_\tau$ gets generated from charged fermion Yukawa couplings on RGE running.  This implies that the natural seesaw scale is $M_N = f v_R \gtrsim 2000 TeV$.  

The region $v_R \leq 50 TeV$, that can be probed by LHC and a future $100 TeV$ collider, requires a smaller Dirac Yukawa coupling and therefore needs to be tuned.    This can be done by adjusting parameters of the Higgs potential so that, on RGE running,  the contribution to $m_D (= h_D v_{wk})$ from the VEV of the first standard model Higgs doublet in the bi-doublet, gets canceled by the contribution from the VEV of the second doublet to a  $10\%$ to $1\%$ level.  

Further for $v_R \leq 50 TeV$, we find that the $h_D$ generated on RGE running increases the mass $m_\nu$ of the third generation light neutrino, and this suppresses the large mixing angle $\tan^2\theta_{23}$ of the PMNS matrix. Thus we lose large neutrino mixing on renormalization running.  By  tuning we can  cancel the contribution to $m_\nu$ at some chosen renormalization scale.   However at just a  few  orders of magnitude lower or higher scales the suppression of $\theta_{23}$ may persist. The problem is more severe for the LHC than it is for the $100 TeV$ collider, where tuning may be more effective for higher values of $M_N \sim 30 TeV$ 

The considerations in this work are quite general and apply also to cases where the discrete $P$ symmetry is broken at a higher scale than $v_R$, and to SUSY LR models.  

Naturalness itself provides a way to test the natural regions of the left-right symmetric model. Since leptonic CP phases generate the strong CP phase in one loop, they must be negligibly small in significant regions of parameter space.   Fig~\ref{fig1} provides a comprehensive picture for testing the minimal left-right symmetric model and shows the natural and fine tuned regions, and the reach of collider and neutrino experiments for seesaw scales from $TeV$ to $10^{15} GeV$.


\begin{acknowledgments} I am grateful to Sonali Tamhankar for helpful discussions and suggesting R programming language for calculations. 
\end{acknowledgments}

\bibliography{natural_bibtex}

\end{document}